\documentclass[aps,reprint,amsmath, amssymb,groupedaddress,floatfix]{revtex4-1}
\usepackage{natbib}
\usepackage{graphicx}
\usepackage{bm}
\usepackage{color,soul}
\usepackage{hyperref}
\usepackage{float}
\usepackage{csquotes}
\usepackage{color,soul}
\usepackage{hyperref}
\hypersetup{colorlinks=true, urlcolor=blue, citecolor=blue}

\begin{document}

\preprint{APS}
\title{Large power dissipation of hot Dirac fermions in twisted bilayer graphene }

\author{S. S. Kubakaddi}
\email{sskubakaddi@gmail.com}
\affiliation{
 Department of Physics, K. L. E. Technological University, Hubballi-580031, Karnataka, India
}

\date{\today}
\begin{abstract}

We have carried out a theoretical investigation of hot electron power loss  $P$, involving electron-acoustic phonon interaction, as a function of twist angle $\theta$, electron temperature $T_e$ and electron density $n_s$ in twisted bilayer graphene (tBLG). It is found that as $\theta$  decreases closer to magic angle $\theta_m$, $P$ enhances strongly and $\theta$ acts as an important  tunable parameter, apart from $T_e$ and $n_s$. In the range of $T_e$ =1-50 K, this enhancement is $\sim$ 250-450 times the $P$ in monolayer graphene (MLG), which is manifestation of the great suppression of Fermi velocity ${v_F}^*$ of electrons in moiré flat band. As $\theta$ increases away from $\theta_m$, the impact of $\theta$ on $P$ decreases, tending to that  of MLG at $\theta$ $\sim$ 3$^{\circ}$.  In the Bloch-Grüneisen (BG) regime, $P$ $\sim$ ${T_e}^4$, ${n_s}^{-1/2}$ and ${v_F}^{*-2}$. In the higher temperature region ($\sim$10- 50 K), $P$ $\sim$ ${T_e}^{\delta}$, with $\delta \sim$ 2.0,  and the behavior is still super linear in $T_e$, unlike the  phonon limited  linear-in-  $T$ ( lattice temperature)  resistivity $\rho_p$. $P$ is weakly, decreasing (increasing) with increasing $n_s$ at lower (higher) $T_e$, as found in MLG. The energy relaxation time $\tau_e$ is also discussed as a function of $\theta$ and $T_e$. Expressing the power loss $P = F_e(T_e)- F_e(T)$, in the BG regime, we have obtained a simple and  useful relation $F_e(T) \mu_p (T) = (e{v_s}^2$/2) i.e. $Fe(T) = (n_se^2 {v_s}^2/2)\rho_p$, where $\mu_p$ is the acoustic phonon limited mobility and $v_s$ is the acoustic phonon velocity. The $\rho_p$ estimated from this relation using our calculated $F_e(T)$ is nearly agreeing with the $\rho_p$ of Wu et al (\textcolor{blue}{Phys. Rev. B 99, 165112 (2019)}).
\end{abstract}
\maketitle

\section{ INTRODUCTION}

Recent pioneering experimental discoveries in twisted bilayer graphene (tBLG) by Cao et al \cite{1, 2}, have created  great interest in the study of their electronic properties and  has ushered in a new era in the condensed matter physics \cite{3, 4, 5, 6, 7, 8, 9, 10, 11, 12, 13}. Among the discoveries, the existence of correlated insulating phases and superconductivity at low temperatures and a highly resistive linear-in-temperature $T$  resistivity $\rho$ at high temperature,  are remarkable and exciting \cite{1, 2,4}. Very recently, the observation of a quantum anomalous Hall effect in twisted bilayer graphene aligned to hexagonal boron nitride has been reported in tBLG \cite{9}. In tBLG a small twist angle $\theta$, near the magic angle $\theta_m$, between the two layers plays the most significant role and acts as one of the tunable parameters, similar to the carrier density $n_s$ and temperature $T$, of the samples in limiting their electronic properties \cite{1, 2, 4, 12, 13}. The transport results of Cao et al  \cite{11}  establish magic angle bilayer graphene as a highly tunable platform to investigate ‘strange metal’ behavior. Because of the twist between the layers the band structure is a moiré flat band with the twist angle dependent suppressed Fermi velocity ${v_F}^*(\theta$) $< v_F$, the bare Fermi velocity in monolayer graphene, and the large density of states $D(E_k)$ near $\theta_m$ at which ${v_F}^*(\theta$) = 0 \cite{12,13,14}. The strongly enhanced electrical resistivity $\rho$, near $\theta_m$, with linear-in-temperature behavior has been observed for $T > \sim$ 5 K \cite{10, 11}.

Theoretically, the electrical resistivity has been investigated in tBLG, at higher temperature and away from the moiré miniband edge, by considering the effect of electron- acoustic phonon (el-ap) interaction \cite{10, 12, 13}. It is shown that the phonon limited resistivity $\rho_p$  = $\rho$ (T, $\theta$) is strongly enhanced in magnitude, twist-angle dependent and linear-in- $T$  occurring for $T > T_L$, where $T_L$ (on the order of few kelvins) is the temperature above which linearity in $\rho (T, \theta)$ develops. This linear-in- $T$ is observed for $T_L = T_{BG}/8$ \cite{13}, where $T_{BG} = 2\hbar v_sk_F/k_B$ is the Bloch- Grüneisen (BG) temperature, $v_s$ is the acoustic phonon velocity, and $k_F =\sqrt{\pi n_s/2}$ is the Fermi wave vector in tBLG. The enhancement in $\rho (T, \theta)$, about three orders of magnitude greater than that in monolayer graphene  (MLG) at $T  \sim$ 10 K, is shown to arise from the large increase in the effective el-ap scattering in tBLG  due to the suppression of  $v_F$ induced by the moiré flat band. In the metallic regime  i.e. for $T > T_m (< T_L)$,  where $T_m$  is the metallic  temperature, above which $d\rho (T, \theta)/ dT > 0$, and it is $n_s$ and $\theta$ dependent. The $\rho (T, \theta$) is found to increase with increasing $T$ as the twist angle $\theta$ approaches $\theta_m$. The linear dispersion taken for the Dirac fermions in tBLG is an approximation that is valid for Fermi energy near the Dirac point and hence its transport study is limited to the $n_s \leq 10^{12}$ cm$^{- 2}$. Interestingly, it is also shown that the same enhanced el-ap interaction can also produce superconductivity with $T_c \sim 1K$ in s, p, d and f orbital pairing channels  \cite {3, 12}.

The theory of Wu et al \cite {12, 13} explains the available  experimental data of $\rho$ well for $T  >$ 5K \cite{10, 11}. In their theory, all the effects of   disorder, impurities and defects  are ignored  assuming that the system is extremely clean and the Fermi energy is  slightly away from the Dirac point. However, the hot electron relaxation is an important transport property which is controlled by only electron-phonon interaction and independent of disorders and impurities. 

The electron system in samples  subject to large electric fields or photoexcitation establishes its internal thermal equilibrium at an electron temperature  $T_e$ greater than the lattice temperature $T$ because electron-electron interaction occurs at the  time scale of several femtoseconds  which is much smaller than the electron-phonon scattering time. Consequently, the electron system is driven out of equilibrium with the lattice. In steady state, these electrons will relax towards equilibrium with the lattice by dissipating energy with phonons as the cooling channels. The study of hot electron power loss $P$ is important as it affects thermal dissipation and heat management which are key issues in nanoscale electronics device. Moreover, it is  crucial for  applications in variety of devices such as calorimeters, bolometers, infrared detectors, ultrafast electronics and high speed communications.  Hot electron cooling  has been extensively studied theoretically and experimentally in MLG  \cite{15, 16, 17, 18, 19, 20, 21, 22, 23, 24} and conventional bilayer graphene (BLG)  \cite{18, 25, 26, 27}. 

In the present work, we investigate the effect of enhanced el-ap coupling on the power dissipation $P$ of the hot electrons in moiré flat band in tBLG. It is studied as a function of twist angle, electron temperature and electron density. We show that the twist angle $\theta$ acts as one of the strong tunable parameters  of $P$. Additionally, a relation between power loss and phonon limited mobility $\mu_p$ is brought out in BG regime.

\section{Theoretical model}

Wu et al \cite{12}  have used the effective Dirac Hamiltonian with a renormalized velocity for electron energy spectrum, in order to obtain their analytical results. In moiré flat band, the electron energy spectrum is assumed to be Dirac dispersion $E_k = \hbar {v_F}^* \vert k \vert$, which  is an approximation that is valid for near Dirac point, with an effective Fermi velocity ${v_F}^* \equiv  {v_F}^*(\theta)$. Because of this approximation  our theory will be limited to the carrier density  $n_s \leq 10^{12}$ cm$^{- 2}$. The density of states is $D(E_k) = g(E_k)/ [2\pi {(\hbar {v_F}^*)}^2]$ with the degeneracy $g$ =  $g_s$ $g_v$ $g_l$, where $g_s$, $g_v$ and $g_l$ are,  respectively, spin, valley and layer degeneracy each with  the value of 2. We consider electron-acoustic phonon interaction within the deformation potential approximation with the longitudinal acoustic (LA)  phonons of energy $\hbar\omega_q$ and wave vector \textbf{q}  interacting with the tBLG Dirac electrons in the moiré miniband. The LA  phonons in tBLG  are assumed to be unaffected by the tBLG structure and are taken to be  the same as the MLG phonons. In MLG the experimental observations of  electrical conductivity \cite{28}  and power loss \cite{19,20,22} are  very well explained by the electron interaction with only LA phonons, without screening. Wu et al \cite{12,13} have explained the linear-in-T resistivity data in tBLG with only electron-LA phonon interaction.  
We use the  modified ordinary  el-ap matrix element \cite{12} ${\vert M(q) \vert}^2= [(D^2 \hbar q F(\theta))/(2A\rho_m v_s)][1-(q^2/4k^2)]$ where $D$ is the first-order acoustic deformation potential coupling constant, $A$ is the area of the tBLG, $\rho_m$ is the areal  mass density and $v_s$ is the LA phonon velocity. The detailed tBLG moiré wave function gives rise to the form factor function $F(\theta)$ which modifies the el-ap interaction matrix element in tBLG as compared with the MLG \cite{12}. It is shown to be between  0.5 and 1.0 and  being nearly parabolic for 1$^\circ$ $<$  $\theta <$ 2$^\circ$ in the neighborhood of a minimum at $\theta$ =  $\sim$ 1.3$^\circ$  \cite{13}.  Following  the Refs. \cite{15, 29, 30}, and taking care of additional layer degeneracy, we obtain an expression for the electron power loss in tBLG  and it is given by

\begin{widetext}

\begin{equation}
    P = - \frac{gD^2F(\theta)}{4\pi^2n_s\rho_m\hbar^5{v_s}^3{{v_F}^*}^3} \int_0^\infty d(\hbar\omega_q) {(\hbar\omega_q)}^2 \int_\gamma^\infty dE_k \frac{(E_k+\hbar\omega_q)}{{[1-{(\gamma/E_k)}^2]}^{1/2}}   \times G(E_q,E_k)[N_q(T_e)-N_q(T)][f(E_k)-f(E_k+\hbar\omega_q)],
\end{equation}
\end{widetext}
where $n_s$ is the electron density, $\gamma = (E_q/2)$, $E_q = \hbar {v_F}^*q$,  $N_q(T) = {[exp(\hbar\omega_q/k_BT) -1]}^{-1}$ is the Bose-Einstein distribution at lattice temperature $T$ and $G(E_q, E_k)  =[1-{(\gamma/E_k)}^2]$, is due to  the  spinor  wave function of the electron in the electron -phonon matrix element, in the quasi-elastic approximation \cite{15}. By setting F($\theta$) =1, $g_l$  = 1 and ${v_F}^*(\theta) = v_F$ in Eq.(1), we regain the equation that is applicable to MLG \cite{15} and silicene \cite{31}, similar to the acoustic phonon induced resistivity in tBLG \cite{10,12}. 
The twist angle dependence of ${v_F}^* \equiv {v_F}^* (\theta)$  is shown to be very well approximated by \cite{10,13} 
\begin{equation}
   {v_F}^* \approx 0.5 \vert \theta-\theta_m \vert v_F, 
\end{equation}
which clearly indicates that twist angle effect is very large for $\theta$ closer to  $\theta_m$. We use this relation while computing $P$ for different twist angles.

In the Bloch-Grüneisen (BG) regime $T$, $T_e$ $<< T_{BG}$,  $q<<$ 2$k_F$, the power loss is given by    
\begin{equation}
   P = \Sigma ({T_e}^4-T^4)/{n_s}^{1/2}, 
\end{equation}
where $\Sigma=\Sigma_0 (D^2/{v_s}^3)$ and $\Sigma_0 = (g\pi^{5/2}{k_B}^4 F(\theta)) / (60\sqrt{2}\rho_m\hbar^4{{v_F}^*}^2)$. Hence, in BG regime $P\sim T^4$, ${n_s}^{-1/2}$ and ${{v_F}^*}^{-2}$.

\section{Results and discussion}

We obtain the following numerical results of power loss in tBLG using the parameters  \cite{12,13}:  $\rho_m = 7.6\times10^{-8}$ gm/cm$^2$, $\theta_m$ = 1.02$^\circ$,   $v_s$ = 2$\times$10$^6$ cm/s, $v_F$ =  1$\times$10$^8$ cm/s  and   $D$ = 20 eV  \cite{15,17,20,28,32,33}, noting that Polshyn et al \cite{10} and Wu et al  \cite{12} have used $D$ = 25 ± 5 eV. In order to bring out the angular dependence of the power loss, we   confine our illustrations for $\theta$ =1.1$^\circ$, 1.2$^\circ$ and 1.3$^\circ$ which are closer to magic angle $\theta_m$ =1.02$^\circ$. For these angles, the effective Fermi velocity ${v_F}^*$ = 4, 9 and  14$\times$10$^6$ cm/s ($>$ 1.5 $v_s$ \cite{12}), respectively, which are much smaller than the bare $v_F$,  and the effect of ${v_F}^*$  on the transport coefficients will be very large. For, further increase of   $\theta$, ${v_F}^*$ tends to $v_F$ at about 3.0$^\circ$.  The values of the function $F(\theta)$ for different $\theta$ are taken from figure 3 of Das Sarma et al \cite{13}, and because of its value between 0.5 and 1, it will  have smaller influence on $P$ than ${v_F}^*$. We have presented the calculations for lattice temperature $T$ = 0.1 K, and $n_s = 0.1 - 1 n_0$, with $n_0$ =1$\times$10$^{12}$ cm$^{-2}$, which keeps us slightly away from the Dirac point and within the linear region of moiré flat band. For $n_s = N n_0$, $T_{BG} = 38.3\sqrt{N}$ which is smaller by a factor of $\sqrt{2}$ compared to MLG.

\begin{figure}
\centering
\includegraphics[angle=0.0,origin=c,height=8.5cm,width=8.5cm]{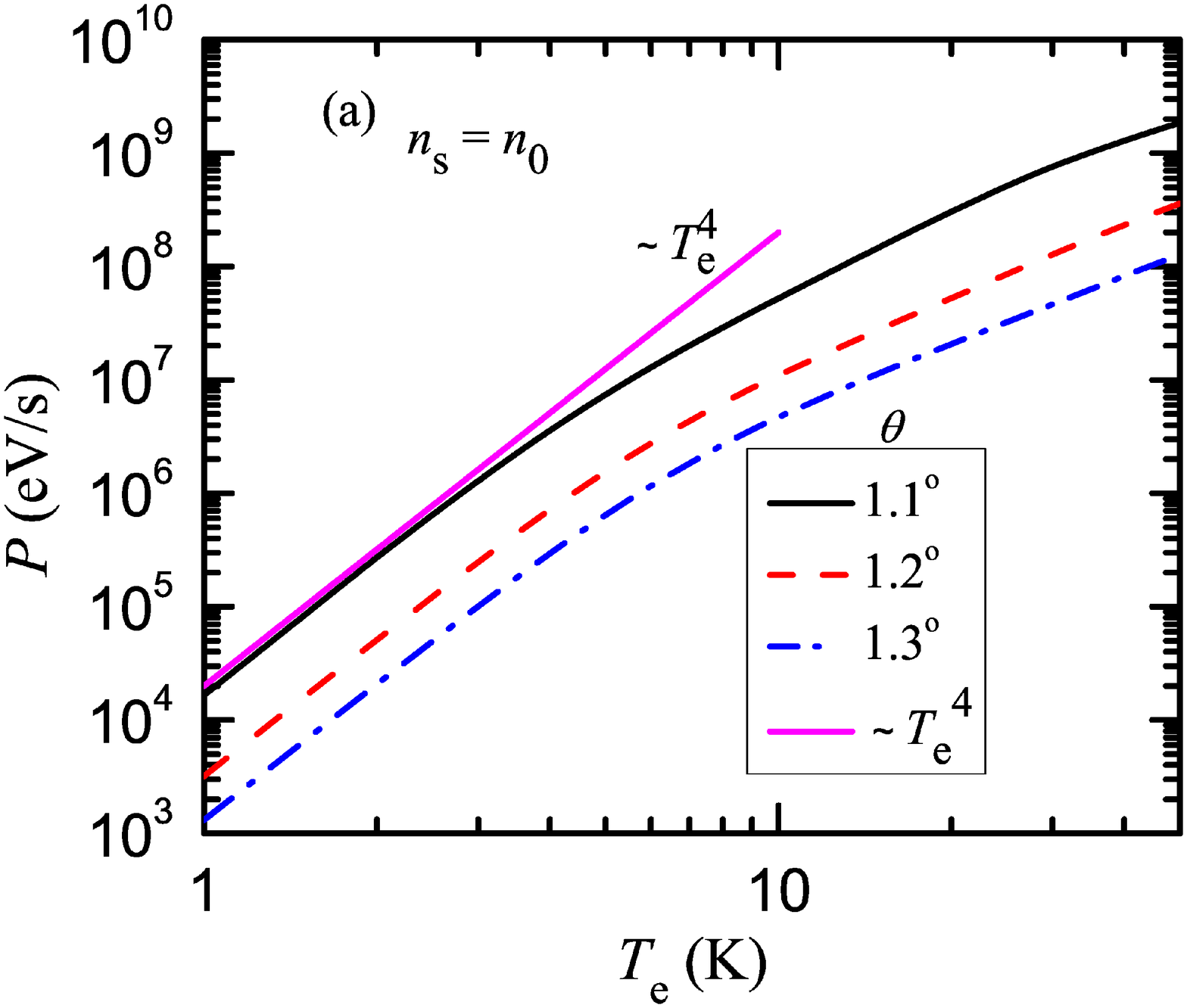}
\end{figure}

\begin{figure}
\centering
\includegraphics[angle=0.0,origin=c,height=8.5cm,width=8.5cm]{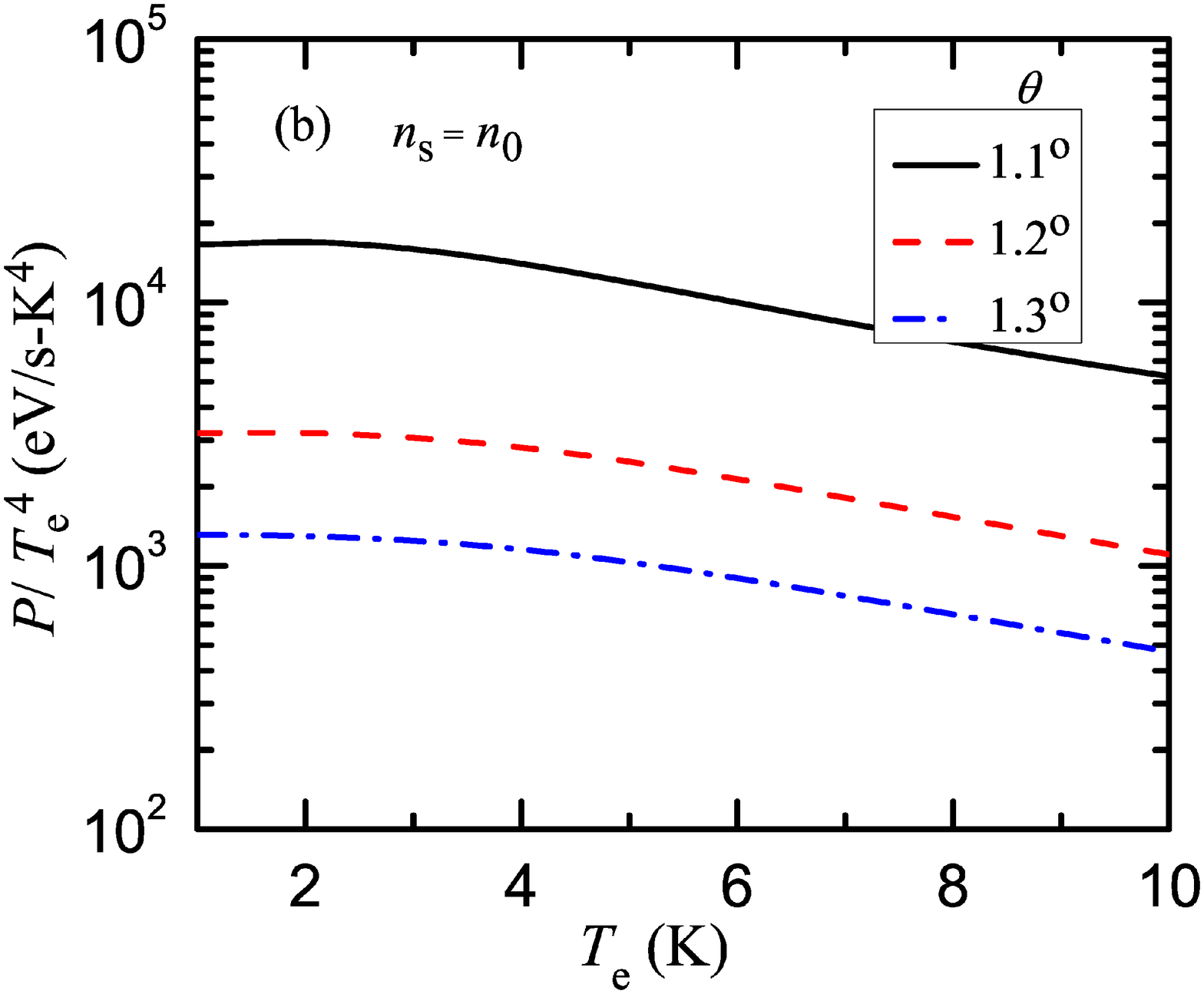}
\end{figure}

\begin{figure}
\centering
\includegraphics[angle=0.0,origin=c,height=8.5cm,width=8.5cm]{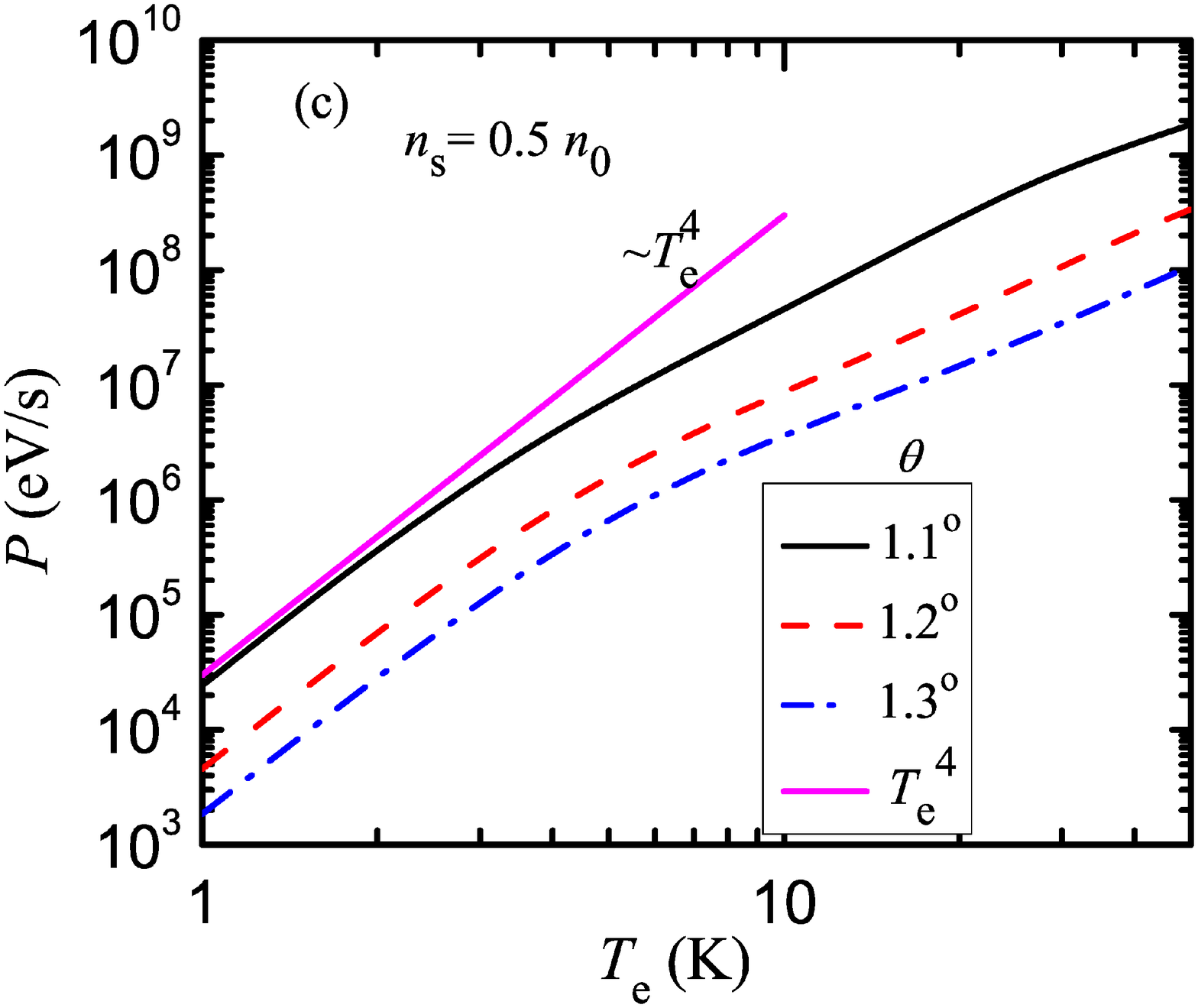}
\caption{  Electron temperature $T_e$  dependence of the power loss $P$ in 
tBLG for twist angle  $\theta$ = 1.1$^{\circ}$, 1.2$^{\circ}$  and 1.3$^{\circ}$. (a) $P$ vs $T_e$  for n$_s$ = n$_0$,  
(b) $P/T_e^4$ vs  $T_e$  for $n_s$ = $n_0$ and (c) $P$  vs $T_e$  for $n_s$=0.5$n_0$. }
\label{fig1}
\end{figure}

\begin{figure}
\centering
\includegraphics[angle=0.0,origin=c,height=8.5cm,width=8.5cm]{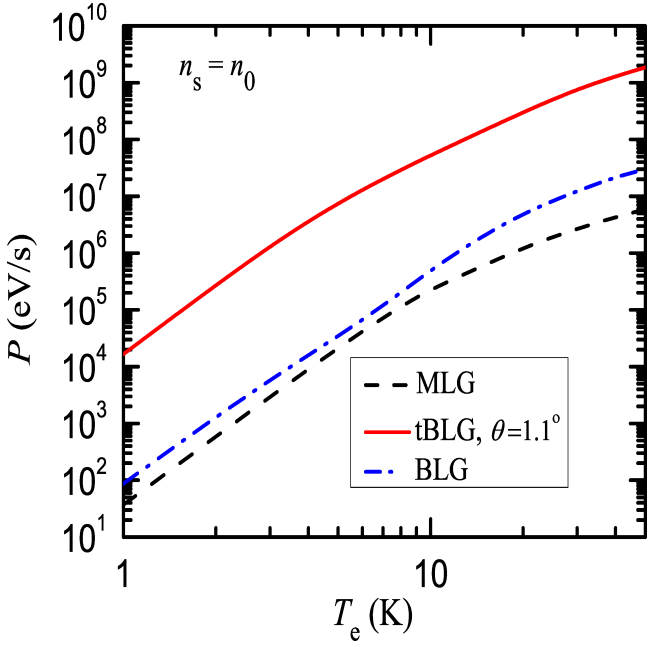}
\caption{ Power loss $P$ as a function of electron temperature $T_e$  for $n_s$ = $n_0$   in  tBLG ( $\theta$ = 1.1$^{\circ}$), MLG and BLG.  }
\label{fig2}
\end{figure}

First we explore the dependence of power loss $P$ on electron temperature $T_e$ for twist angles $\theta$ =1.1$^\circ$, 1.2$^\circ$ and 1.3$^\circ$. In figure \ref{fig1}a, $P$ is presented as a function of $T_e$ (1-50 K) for $n_s = n_0$. For all the $\theta$, we observe the generic nature of the behavior, where in at very low $T_e$ power loss increases rapidly then slows down at higher temperature. For the temperatures  $T_e$ $<<$ $T_{BG}$, the rapid increase may be attributed to the increasing number of phonons as their wave vector $q \approx  k_B T_e/\hbar v_s$ increases linearly with $T_e$. For $\theta$ =1.1$^\circ$, the  power law $P \sim {T_e}^4$ is found to be obeyed for $T_e < \sim$ 2.5 K, which is about $T_{BG}$/15. The exponent 4 of $T_e$ is manifestation of two-dimensional phonons with unscreened electron-phonon coupling. In order to see the effect of $\theta$ on the range of validity of  the power law, we have plotted $P / {T_e}^4$ vs $T_e$ in figure \ref{fig1}b. It is found that, as $\theta$ increases the range of $T_e$ in which power law is obeyed marginally increases. For example, for $\theta$ =1.2$^\circ$ and 1.3$^\circ$, power law is found to be  satisfied for $T_e$ up to about 3 and 3.5 K, respectively, although $T_{BG}$ is same. This happens   because as $\theta$ increases ${v_F}^*$ also increases and tends towards $v_F$. In the BG regime, in which $P \sim$ ${{v_F}^*}^{ -2}$, we find  $\Sigma$ = 2.66$\times$10$^{-15}/\sqrt{N}$ W/K$^4$-cm,  5.13$\times$10$^{-16}$/$\sqrt{N}$ W/K$^4$-cm and 2.1$\times$10$^{-16}$ /$\sqrt{N}$  W/K$^4$, for $\theta$ =1.1$^\circ$, 1.2$^\circ$  and 1.3$^\circ$, respectively, as compared to 5.23$\times10^{-18}/\sqrt{N}$ W/K$^4$-cm in MLG. In the higher temperature region of $T_e$ = 10 $-$ 50 K (30 $-$ 50 K), $P \sim {T_e}^\delta$  with  $\delta$ $\sim$ 2.0 $-$ 2.2 ($\sim$ 1.7 $-$ 2.0), for all $\theta$s, as compared to the resistivity which is found to show linear-in-temperature for  temperature $\geq$ $T_{BG}$/8 \cite{13}.

In figure \ref{fig1}c, $T_e$ dependence of $P$ is shown for $n_s$ = 0.5 $n_0$ and the same behavior is observed as in figure \ref{fig1}a. However, for the same $\theta$, in the low temperature region $P$ is found to be marginally larger (smaller) at lower (higher) $T_e$ than that for $n_s$ = $n_0$. We observe that the  temperature below which the  power law $P \sim {T_e}^4$ is obeyed shifts to lower side for smaller $n_s$. For example, for $\theta$ =1.1$^\circ$ power law is obeyed for $T_e < \sim$2 K, which may be attributed to the lower $T_{BG}$ = 27.0 K. More importantly, from the  figures \ref{fig1}a and \ref{fig1}c, we find that $\theta$ acts as a tunable parameter of power loss, in addition to $T_e$ and $n_s$. The influence of $\theta$ on $P$ is  very much large compared to $n_s$. For $\theta$ =1.1$^\circ$ and 1.3$^\circ$, for the $T_e$ range considered, $P$ is in the range $\sim 10^4-10^9$ eV/s and $\sim 10^3-10^8$ eV/s, respectively. These values are comparable to those in monolayer MoS$_2$ \cite{30} but about three and four orders  of magnitude greater than those in GaAs heterojunction \cite{34} and Si-inversion layer \cite{35}, respectively. 

In order to compare the power loss in tBLG with that in MLG and conventional BLG, $P$ dependence on $T_e$ is depicted in figure \ref{fig2}, for $n_s$ = $n_0$ with $P$(tBLG) taken for $\theta$ =1.1$^\circ$. We find that the power loss in tBLG is very large ($\sim 2\times10^4 - 2\times 10^9$ eV/s) compared to that in MLG ($\sim 4\times 10^1 - 6\times10^6$ eV/s) and BLG ($\sim 9\times10^1 - 3\times10^7$ eV/s) \cite{26}. Defining a ratio $R_p$ = $P$(tBLG)/ $P$(MLG), it is found that $R_p$ = $\sim$ 450, 260 and 300, respectively, at 1, 10 and 50 K, and $R_p$ is expected to be smaller for larger $\theta$. This enhancement is attributed to the significantly reduced ${v_F}^*$. This may be compared with the $\rho$ enhancement in tBLG,  which is of  three orders of magnitude greater than that in MLG at $\sim$10 K \cite{12,13}.  It is also noticed that the range of $T_e$ in which power law is obeyed is much larger in MLG than in tBLG. 

\begin{figure}
\centering
\includegraphics[angle=0.0,origin=c,height=8.5cm,width=8.5cm]{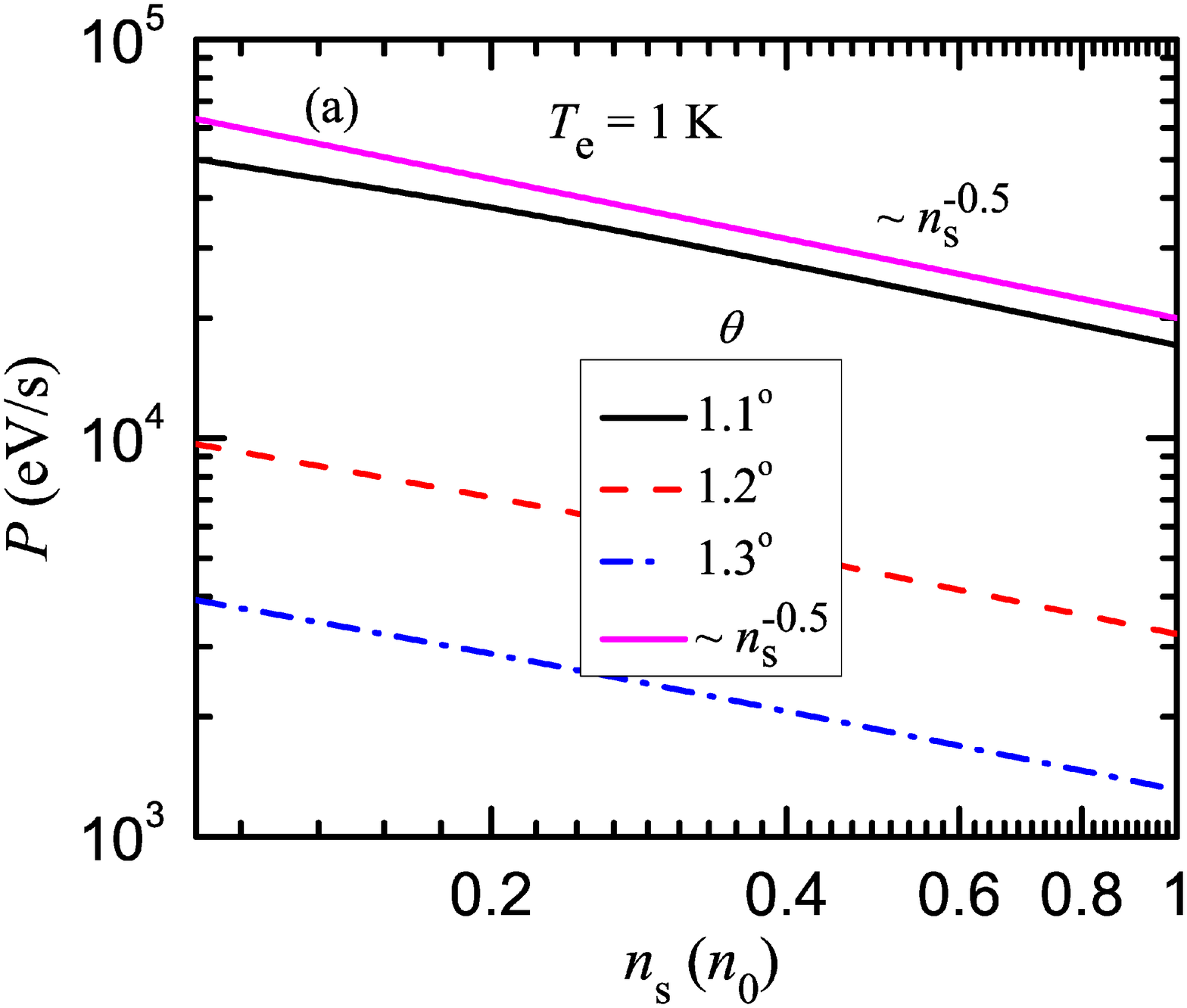}
\end{figure}
\begin{figure}
\centering
\includegraphics[angle=0.0,origin=c,height=8.5cm,width=8.5cm]{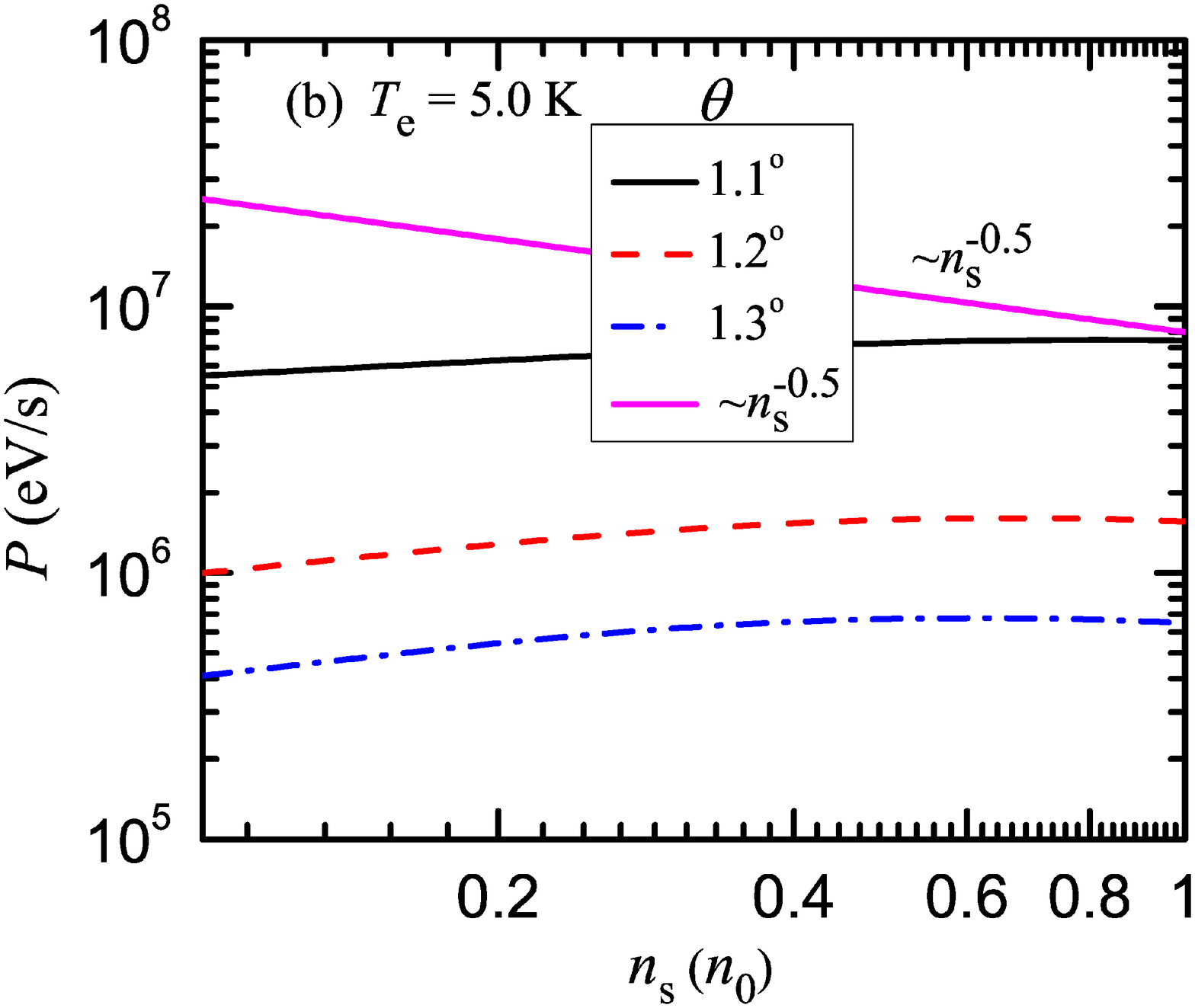}
\caption{ Power loss $P$ as a function of electron density $n_s$   in  tBLG for $\theta$ = 1.1$^{\circ}$, 1.2$^{\circ}$  and 1.3$^{\circ}$. (a) $T_e$ = 1 K and (b) $T_e$ = 5K. }
\label{fig3}
\end{figure}

We have presented in figure \ref{fig3} the electron density (=0.1-1.0 $n_0$) dependence of the power dissipation for two electron temperatures $T_e$ = 1 K (figure \ref{fig3}a) and 5 K (figure \ref{fig3}b). For $n_s$ = 0.1(1.0) $n_0$  the $T_{BG}$ =  12.1 (38.3) K. From figure \ref{fig3}a, we see that $P$ is found to decrease with increasing  $n_s$, as found in MLG \cite{15,22}, with power law $P$ $\sim$ ${n_s}^{-1/2}$ being followed at larger $n_s$ and small deviation occurring at lower $n_s$. This is due to the fact that $T_{BG}$ goes on decreasing with decreasing $n_s$. The $P$ $\sim  {n_s}^{-1/2}$ dependence in tBLG is in contrast to the $P$ $\sim {n_s}^{-3/2}$ dependence in conventional BLG \cite{26,27}.  On the other hand for $T_e$ = 5 K (figure \ref{fig3}b), $P$ increases (flattens) with increasing $n_s$ in the low (high) $n_s$ region, because we are moving away from the $T_e$ $<<$ $T_{BG}$ region. 
\begin{figure}[h]
\centering
\includegraphics[angle=0.0,origin=c,height=8.5cm,width=8.5cm]{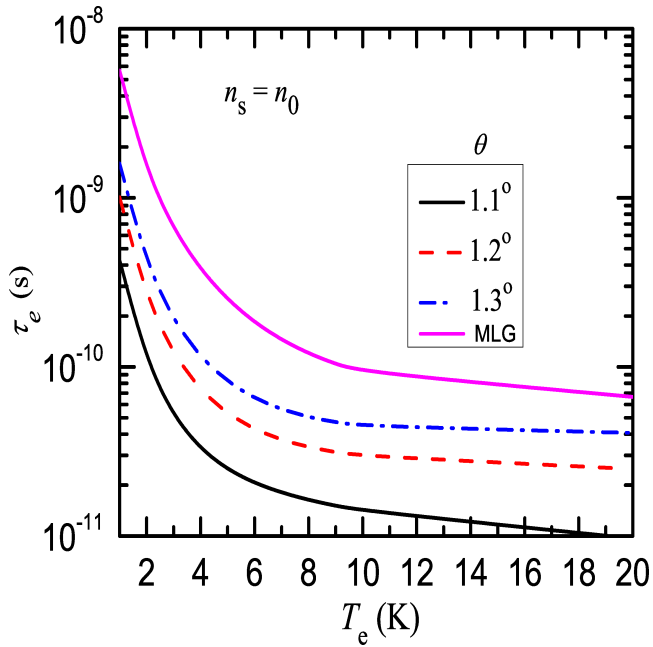}
\caption{  Energy  relaxation  time  $\tau_e$   as  a  function  of  electron temperature $T_e$ for $n_s$ =  $n_0$ in tBLG ($\theta$ = 1.1$^{\circ}$, 1.2$^{\circ}$  and 1.3$^{\circ}$) and 
MLG. }
\label{fig4}
\end{figure}

The energy relaxation time $\tau_e$ is another important quantity studied in the hot electron relaxation process, as it determines the samples suitability for its applications in optical detectors (bolometer, calorimeter and infrared detectors) and .high speed devices. For a degenerate electron gas it is given by $\tau_e$ = $[(p+1) {(\pi k_B)}^2 ({T_e}^2 -  T^2) / (6 E_FP)]$, where $p$ is the exponent of energy in density of states and $E_F$ is the Fermi energy \cite{20,36}. In BG regime, since $P$ $\sim {T_e}^4$ and ${n_s}^{-1/2}$, we find  $\tau_e$ $\sim {T_e}^{-2}$ and independent of $n_s$ (as $E_F$ $\sim$ ${n_s}^{1/2}$). In figure \ref{fig4}, $\tau_e$ is presented as a function of $T_e$, for $\theta$ = 1.1$^\circ$, 1.2$^\circ$ and 1.3$^\circ$, in tBLG along with the $\tau_e$ in MLG for $n_s$ = $n_0$. In both tBLG and MLG, $\tau_e$ is found to decrease with increasing  $T_e$ and the decrease is rapid at lower temperature ($< \sim$10 K). It is found that $\tau_e$ in tBLG, for $\theta$ =1.1$^\circ$, is an order of magnitude smaller than that in MLG and this difference decreases with increasing $\theta$.  The ratio $\tau_e$ (MLG)/ $\tau_e$ (tBLG), for $\theta$ =1.1$^\circ$, is found to be  10.0, 6.6 and 6.9,  respectively, for $T_e$ = 5, 10 and 20 K. This ratio is not as large as the ratio of $P$’s, because of the product $E_FP$ in the denominator of the expression for $\tau_e$, noting that for the same $n_s$, the $E_F$ in tBLG is much smaller than that in MLG. By increasing $\theta$ the $\tau_e$ increases significantly, indicating that twist angle is an important tunable parameter for $\tau_e$ also. It may be noted that samples with faster energy relaxation (i.e. smaller $\tau_e$) find applications in ultrafast electronics and high speed communications. On the other hand, samples with longer energy relaxation time are preferred in photodetectors and energy harvesting devices like hot carrier solar cells.

Finally, in BG regime, we bring out a simple relation of $P$ with phonon limited mobility $\mu_p$ and resistivity $\rho_p$ in tBLG. In this regime,  $P$, $\mu_p$ and $\rho_p$ are sensitive measures of the el-ap coupling. While $P$ is determined by the energy relaxation through el-ap interaction, $\mu_p$ and $\rho_p$ involve momentum relaxation through the same mechanism. A relation between these measurable properties is expected because of the same underlying mechanism. This kind of relation between $P$ and  $\mu_p$ is listed for different electron systems in Ref \cite{37}. In tBLG, the equation $\rho_p (T) = A T^4$ for the phonon limited resistivity is obtained from Min et al \cite{38} with suitable replacement of $g_s g_v$ by $g_s g_v g_l$, $k_F = \sqrt{(\pi n_s/2)}$ and inserting $F(\theta)$ in the numerator in their Eq. (8)  for A. There by, using the relation $\mu_p (T) = 1/(n_s e \rho_p(T))$, the  phonon limited mobility is found to be $\mu_p (T) = [15(ge\hbar^4\rho_m {v_s}^5 {{v_F}^*}^2{n_s}^{1/2}T^{-4})][16\sqrt{2}\pi^{5/2}D^2{k_B}^4F(\theta)]$, where $e$ is the electron charge. Expressing Eq. (3) as $P = F_e (T_e)- F_e (T)$  \cite{15,34}, where $F_e (T) = \Sigma T^4/ {n_s}^{1/2}$ and $P = F_e (T_e)$ for $T_e >> T$,  we  obtain a very simple relation $F_e (T) \mu_p (T) = (e{v_s}^2/2)$, which is exactly same as that of MLG \cite{37}. This relation is  analogous to Herring’s law \cite{39}, which relates  phonon-drag thermopower $S^g$ and $\mu_p$. Alternatively, power loss can be related to $\rho_p$ by the formula $F_e (T) = (n_se^2{v_s}^2/2) \rho_p (T)$. The advantage of these relations  is, if $F_e (T)$ is measured then $\mu_p (T)$ and $\rho_p (T)$ can be determined or the vice-versa, and the measurements of power loss may be preferred as it is independent of lattice disorders and impurities. From our calculated value of  $P  = F_e (T)$  = 4.45$\times10^{-14}$ W at 2 K for $\theta$ =1.1$^\circ$ and $n_s$ = $n_0$, we estimate $\rho_p (T)$ = 0.8 $\Omega$,  which is nearly agreeing with  the value obtained by Wu et al (see figure 4a of \cite{12}), and $\mu_p(T)$ = 7.5$\times 10^6$ cm$^2$/V$-$s.

We would like to make the following remarks. In the literature the values of $\theta_m$ given are varying between 1.02$^\circ$ to 1.1$^\circ$  \cite{1,8,12,13}. However, we believe that our findings and analysis with $\theta_m$ = 1.02$^\circ$  \cite{12,13} hold good for the $\theta$ values closer to any chosen $\theta_m$. We want to emphasize that, our analytical results will be of great help to experimental researchers and secondly can be used to determine $D$ as the measurements of $P$ are independent of lattice disorder and impurities, unlike resistivity.

\section{Conclusions}

We have studied the hot electron power loss $P$ due to the simple  acoustic phonon interaction, via deformation potential coupling, in tBLG of low electron density   $n_s \leq$ 10$^{12}$ cm$^{-2}$ for small twist angles $\theta$  and for $T_e$ $\geq$ 1 K. For $\theta$ closer to the magic angle, $P$ is enhanced by a few hundred times that in MLG due to the great suppression of the Fermi velocity ${v_F}^*$ leading to the strong el-ap scattering. Consequently, twist angle emerges as an additional important tunable parameter of $P$. Although BG regime power law $P \sim {T_e}^4$ is obeyed in low $T_e$ region, $P$ vs $T_e$ behavior still remains super linear at higher $T_e$ where acoustic phonon limited resistivity $\rho_p$ is linear-in-temperature.  For a given $n_s$, although $T_{BG}$ is independent of  $\theta$, the range of $T_e$ in which $P \sim {T_e}^4$ is obeyed increases marginally with increasing $\theta$. The energy relaxation time $\tau_e$, is found to be smaller by an order of magnitude than in MLG and decreasing with increasing $T_e$. As  $\theta$ approaches $\theta_m$  the $\tau_e$  decreases significantly indicating that $\theta$ can be used as an important parameter to tune $\tau_e$ also. Finally,  simple and useful relations of $P$   with $\mu_p$ and $\rho_p$ are obtained in the BG regime. From the relation between $P$ and $\rho_p$, using our calculated $P$, the estimated value is closer to  the $\rho_p$ of Wu et al \cite{12}.  Experimental observations may test the validity of our predictions.

\bibliography{paper}%
\providecommand{\noopsort}[1]{}\providecommand{\singleletter}[1]{#1}%
\end{document}